\documentstyle[12pt]{article}
\begin{document}
\begin{flushright}
DO-TH-95/05\\
NIIG-DP-95-1\\
April, 1995
\end{flushright}
\begin{center}
{\Large Baryon and Lepton Number Assignment in $E_6$ Models}\\
\vskip .75in

{\large Emmanuel A. Paschos, Utpal Sarkar\footnote{
{\it On sabbatical leave from} Theory Group, Physical Research Laboratory,
Ahmedabad - 380009, India.} and Hiroto So\footnote{{\it On leave from}
Physics Department, Niigata University, Niigata 950-21, Japan}}\\
\vskip .2in
{\it Institut f\"{u}r Physik},\\
{\it Universit\"{a}t Dortmund},\\
{\it D-44221 Dortmund, Germany.}\\

\vskip .5in
\end{center}
\begin{abstract}
\baselineskip 16pt

In $E_6$ models there are new particles whose baryon number is not
uniquely assigned. We point out that the baryon and lepton number
assignment to
these particles can change the baryogenesis scenario significantly.
We consider left-right symmetric extension of the standard model
in which $(B-L)$ quantum number is gauged. The identification of
$(B-L)$ with a generator of $E_6$ is used to define
the baryon and lepton numbers for the exotic particles in a way that the
electroweak baryon and lepton number anomaly corresponding to the
$SU(2)_L$ group vanishes, {\it i.e.}, there is no non-perturbative
baryon or lepton number violation during the electroweak
phase transition. We study some consequences of the new
assignment.

\end{abstract}
\setcounter{footnote}{0}

\newpage
\baselineskip 18pt

In the standard model and its generalizations to Grand Unified
Theories (GUTs) the quantum number $(B-L)$ is conserved \cite{gutrev}.
In fact,
it has been proven that the gauge bosons of GUTs can not decay to
light fermions, which are classified in
$SU(3)_c \times SU(2)_L \times U(1)_Y  [\equiv
G_{std}]$ group, by violating the $(B-L)$ quantum number \cite{wein}.
In addition, it is known that
the contributions of the $SU(2)_L$ anomaly to the baryon and lepton
current satisfies \cite{hooft},
\begin{eqnarray}
\partial j_B &\sim& N_g (3 B_q + B_l) {\rm Tr} F \widetilde{F}
\nonumber \\
\partial j_L &\sim& N_g (3 L_q + L_l) {\rm Tr} F \widetilde{F}
\nonumber
\end{eqnarray}
where, $B_q = 1/3$ and $B_l = 0$ are baryon numbers and
$L_q = 0$ and $L_l = 1$ are lepton numbers
for quarks and leptons, respectively and $N_g$ the number of generations.
This implies,
\begin{equation}
\partial (j_B - j_L) = \partial (j_{(B-L)}) = 0
\end{equation}
The aim of this paper is to investigate the assignment of quantum
numbers of the new (exotic) fermions, which appear in the group
$E_6$ \cite{gutrev,e6rev,us1,us2} and study the anomalies generated
for the $B$ and $L$ currents.
We are interested to produce at a high energy scale (cosmological)
$B$ and $L$ asymmetries and investigate their development to low
temperature, that is the present epoch of the universe.
For the sake of clarity, we present a general discussion
on the origin and conservation of the $B$ and $L$ quantum numbers.

We begin discussing the origin of the $(B-L)$ quantum number.
The $B$ and $L$ quantum numbers are globally conserved in the
Standard Model and they prevent proton decay. This follows from the
internal symmetry and Lorentz invariance of the theory. $B$ and $L$ are
not conserved locally because $\psi_L$ and $\psi_R$ transform
differently. In higher groups like $SU(2)_L \times SU(2)_R \times
U(1)_{B-L} [\equiv G_{LR}]$ \cite{lr} or $SO(10)$ \cite{gutrev},
we can treat left- and right-handed
particles on equal footing and there is a new generator of the
enlarged group corresponding to $(B-L)$.

This points to ways of breaking $B$ and $L$. In the standard
model we can break the global symmetries or a linear combination of them
spontaneously, which implies a Goldstone boson. This is unphysical.
The second method is breaking the local symmetry spontaneously.

In extensions of the standard model $G_{std}$ or the
left-right symmetric model $G_{LR}$, $(B+L)$ is explicitly
broken. The breaking manifests itself as four-fermion operators
composed of quarks and leptons which have dimension six. An
enumeration of the operators reveals \cite{wein} that all
dimension -- 6 $B-$violating operators
which can be constructed from the known quarks and leptons
conserving $G_{std}$ and Lorentz invariance have $B-L =0$ and
$B+L =2$. Operators with $(B-L) \neq 0$ have higher dimension
and are suppressed by factors  $(M_W/M_{GUT})$.

A second contribution to the violation of baryon number is
through $SU(2)_L$ anomalies,
discussed at the beginning of this paper. The anomalies break $(B+L)$
and again preserve $(B-L)$. Thus the anomalies can wash out any $(B+L)$
asymmetry discussed \cite{krs} but will preserve a $(B-L)$ asymmetry with the
possible transformation of $\Delta (B- L) \to \Delta B$ \cite{ht}.

The above discussions indicates that it is difficult to
create a $(B-L)$ asymmetry through couplings of the standard fermions,
because
\begin{itemize}
\item[1.] there are no lowest dimension operators with $B-L \neq 0$
and higher order operators are suppressed by $M_W/M_{GUT}$
\item[2.] there is no $(B-L)$ anomaly which will generate or wash out a
$(B-L)$ asymmetry.
\end{itemize}
These considerations motivated us to investigate the assignment
of quantum numbers $B$ and $L$ to fermions in the group
$E_6$ which contains new fermions.

In $E_6$ GUT, all the chiral fermions belong to the fundamental {\bf 27}
representation.  They are presented in
table 1
\begin{table}[tbp]
\begin{center}
\caption{Quantum numbers of the fermions in terms of the subgroup
$SU(3)_c \otimes SU(2)_L \otimes SU(2)_R \otimes U(1)_{Y_L}
\otimes U(1)_{Y_R}$ of $E_6$}
\vskip .2in
\begin{tabular}{||c|c|cccccc||}
\hline \hline
\multicolumn{2}{||c|}{}&&&&&&\\
\multicolumn{2}{||c|}{Left-handed
fermions}&$SU(3)_c$&$I_{3L}$&$I_{3R}$&$Y_{L}$&$Y_{R}$&$Q$\\
\multicolumn{2}{||c|}{}&&&&&&\\
\hline
&&&&&&&\\
&$Q = \pmatrix{u \cr d}$&3& $\pmatrix{\frac{1}{2} \cr - \frac{1}{2}}$
& 0 & $\frac{1}{3}$ &0 &  $\pmatrix{\frac{2}{3} \cr - \frac{1}{3}}$ \\
&&&&&&&\\
&$u^c$ &$ \bar{3}$& 0 & $- \frac{1}{2}$ & 0 & $- \frac{1}{3}$ &
$- \frac{2}{3}$ \\ &&&&&&&\\
&$d^c$ &$ \bar{3}$& 0 & $ \frac{1}{2}$ &
0 & $- \frac{1}{3}$ & $ \frac{1}{3}$ \\
{\bf STANDARD} &&&&&&&\\
{\bf FERMIONS}&$L_1 = \pmatrix{\nu \cr e}$&1& $\pmatrix{\frac{1}{2} \cr -
\frac{1}{2}}$
& 0 & $- \frac{1}{3}$ &$- \frac{2}{3}$ &
$\pmatrix{0 \cr -1}$ \\
&&&&&&&\\
&$e^c$ &1& 0 & $ \frac{1}{2}$ & $\frac{2}{3}$ & $ \frac{1}{3}$ &
1\\
&&&&&&&\\
&$\nu^c$ &1& 0& $- \frac{1}{2}$ & $\frac{2}{3}$ & $ \frac{1}{3}$ &
0\\
&&&&&&&\\
\hline
\hline
&&&&&&&\\
&$D_1$ & 3& 0 & 0 &  $- \frac{2}{3}$ & 0 & $- \frac{1}{3}$ \\
&&&&&&&\\
&$D_2$ &$ \bar{3}$& 0 & 0 & 0 & $ \frac{2}{3}$ & $ \frac{1}{3}$ \\
&&&&&&&\\
{\bf EXOTIC}&$L_2 = \pmatrix{N_1 \cr E_1}$&1& $\pmatrix{\frac{1}{2} \cr -
\frac{1}{2}}$
&$- \frac{1}{2}$  & $- \frac{1}{3}$ &$ \frac{1}{3}$ &
$\pmatrix{0 \cr -1}$ \\
{\bf FERMIONS}&&&&&&&\\
&$L_3 = \pmatrix{E_2 \cr N_2}$&1& $\pmatrix{\frac{1}{2} \cr - \frac{1}{2}}$
& $ \frac{1}{2}$ & $- \frac{1}{3}$ &$ \frac{1}{3}$ &
$\pmatrix{1 \cr 0}$ \\
&&&&&&&\\
&$L_s$ & 1& 0 & 0 &  $ \frac{2}{3}$ & $- \frac{2}{3}$ & 0 \\
&&&&&&&\\
\hline \hline
\end{tabular}
\end{center}
\end{table} with the quantum number for the diagonal generators.
The electric charge is defined in terms of the $E_6$ generators
as, $$  Q = T_{3L} + \frac{Y}{2} = T_{3L} + T_{3R} +
\frac{Y_L + Y_R}{2} . $$
For the fermions present in the standard model, the baryon and lepton
number assignments are known from experiment. In the notation of
table 1:
\begin{center}
[$B={1 \over 3}$, $L=0$] for (${\bf Q, \overline{u^c},
\overline{d^c}}$) and [$B=0$, $L=1$] for (${\bf L_1, \overline{e^c},
\overline{\nu^c}}$).
\end{center}
Baryon and lepton numbers are not uniquely defined
for the exotic fermions $D_1$, $D_2$, $N_1$, $E_1$, $N_2$,
$E_2$ and $L_s$ and we shall assign them in a systematic manner.
The Yukawa part of the lagrangian invariant under $E_6$ is,
\begin{eqnarray}
{\cal L} &=&  \lambda_{uu} Q^T u^c H_3 + \lambda_{dd} Q^T d^c H_2
+ \lambda_{dD} Q^T D_2 H_1 +  \lambda_{DD} D_1^T D_2 S_2 \nonumber \\
&&+ \lambda_{Dd} D_1^T d^c S_1 + \lambda_{Ee} L_2^T e^c H_1
+ \lambda_{ee} L_1^T e^c H_2 +  \lambda_{EE} L_3^T L_2 S_2
+ \lambda_{eE} L_1^T L_3 S_1 \nonumber \\
&&+  \lambda_{1} L_2^T L_s H_3
+ \lambda_{2} L_3^T L_s H_2 +  \lambda_{3} L_3^T \nu^c H_1
+ \lambda_{4}  L_1^T \nu^c H_3  + h.c.
\end{eqnarray}
We used the convention, $\overline{\psi^c} = \psi^T {\cal C}$
and have not written the ${\cal C}$ operator explicitly.

In writing down this lagrangian we have considered the minimal $E_6$
model, which might have its origin in the superstring theory. All
higgs scalars belong to the fundamental {\bf 27}-representation of $E_6$.
For studying the
low energy effective theory for the baryon nonconservation, we write down
only the higgs fields which can acquire {\it vev}s. We list them in
table 2. The $SU(2)_L$ doublets are written as $H_i$,
while the singlets as $S_i$.

\begin{table}[htb]
\begin{center}
\caption{Quantum numbers of the higgs scalars in terms of the subgroup
$SU(3)_c \otimes SU(2)_L \otimes SU(2)_R
\otimes U(1)_{Y_L} \otimes U(1)_{Y_R}$ of $E_6$}
\vskip .2in
\begin{tabular}{||ccccccc||}
\hline \hline
&&&&&&\\
Higgs&$SU(3)_c$&$I_{3L}$&$I_{3R}$&$Y_{L}$&$Y_{R}$&$Q$\\
&&&&&&\\
\hline
&&&&&&\\
$H_1 $&1& $\pmatrix{ \frac{1}{2} \cr -\frac{1}{2}}$
& 0 & $- \frac{1}{3}$ &$- \frac{2}{3}$ &
$\pmatrix{0 \cr -1}$ \\
&&&&&&\\
$H_2 $&1& $\pmatrix{ \frac{1}{2} \cr -\frac{1}{2}}$
&$- \frac{1}{2}$  & $- \frac{1}{3}$ &$ \frac{1}{3}$ &
$\pmatrix{0 \cr -1}$ \\
&&&&&&\\
$H_3 $&1& $\pmatrix{\frac{1}{2} \cr -\frac{1}{2}}$
& $ \frac{1}{2}$ & $- \frac{1}{3}$ &$ \frac{1}{3}$ &
$\pmatrix{1 \cr 0}$ \\
&&&&&&\\
$S_1$ &1& 0& $- \frac{1}{2}$ & $\frac{2}{3}$ & $ \frac{1}{3}$ &
0\\
&&&&&&\\
$S_2$ & 1& 0 & 0 &  $ \frac{2}{3}$ & $- \frac{2}{3}$ & 0 \\
&&&&&&\\
\hline \hline
\end{tabular}
\end{center}
\end{table}

The first term contributes to the up-type quarks mass. The next four
terms give the mass matrices for the charge (-1/3) particles, {\it i.e.,}
the down-type quarks and the exotic charge (-1/3) color triplets.
The next four terms will
contribute to the charged lepton mass matrices, {\it i.e.,} the usual charged
leptons and the exotic charge (-1) color singlet particles. The eight
and the ninth terms, along with the rest of the terms will contribute
to the mass matrices of the neutral fermions.
The baryon and the lepton numbers of the scalars $H_2$ and $H_3$
are determined through their Yukawa couplings with the usual fermions
to be $B=0$ and $L=0$.

We now extend our knowledge of $B - L$ quantum numbers for the
standard to the exotic fermions. We demand that $B - L$ is a local symmetry
in the left-right extended theory for the exotic fermions
of $E_6$ as well. This is equivalent to saying
that $G_{LR} \subset E_6$ or $SO(10) \subset E_6$ (which is usually
done). This means that we relate the $B$ and $L$ quantum numbers
and the maximal subgroup of $E_6$ for the exotic fermions in the
same way as it is for the standard fermions,
\begin{equation}
{(B-L)} = Y_L + Y_R .  \label{bl}
\end{equation}
In this case $(B-L)$ is a gauge symmetry and the interesting
features of the $SO(10)$ model arising from the
gauged $(B-L)$ symmetry directly follow.

The identification of the $(B-L)$ quantum number with the generators
of $E_6$, eq. (\ref{bl}),
does not give unique values for the baryon and lepton numbers
of the exotic particles. We exploit this freedom and choose the
$B$ and $L$ quantum numbers of the new particles such that the sum
total of contributions to $(B+L)$, from standard and exotic particles,
add up to zero.
Then the extension of the standard model is $SU(2)_L$
anomaly free\footnote{
The absence of the $(B+L)$ anomaly has another consequence that
any primordial lepton asymmetry will not
be changed into the baryon asymmetry. Hence the bounds on the
R-parity violating ($L$--violating) couplings \cite{camp}
does not hold.}.

$D_1$ and $D_2$ are singlets under $SU(2)_L$, they
cannot contribute to the anomaly, which does not help us in
assigning them $B$ and $L$ number. So we impose
one more logical constraint that these particles do not carry
lepton number. Similarly, for the singlet particle $L_s$ we
impose that it does not carry baryon number.

We now turn to the $SU(2)$ anomaly
and find that as long as the sum of the baryon numbers of $L_2$ and
$L_3$ is $-1$, the global $B$ anomaly corresponding to the group
$SU(2)_L$ is zero. Similarly for vanishing lepton anomaly the lepton
numbers of $L_2$ and $L_3$ should add up to $-1$.
Then demanding the validity of eq. (\ref{bl}),
we fix the baryon and lepton numbers of these particles.

The baryon and lepton number anomaly corresponding to the
group $SU(2)_L$ can mediate baryon number violation only through
non-perturbative effects during the electroweak phase transition, which
vanishes for the following choice of baryon and lepton numbers for the
exotic particles,
\begin{center}
$[B=-{2 \over 3};\: L=0]$  for  $({\bf D_1,
\overline{D_2}})$ \hskip .5in $[B=0; \: L=0]$  for
(${\bf L_s}$)\\ and \hskip .5in
$[B=-{1 \over 2}; \:\: L=-{1 \over 2}]$
for $({\bf L_2, L_3})$\\
\end{center}

In general, the couplings of the gauge bosons to fermions produce
change for $\Delta B$ and $\Delta L$ at each vertex.
For vertices where both standard and exotic fermions couple to
the same gauge bosons we may ask if the changes for $\Delta B$ and
$\Delta L$ are the same or different. This question is relevant for
the couplings of (X,Y) and (X',Y') bosons in the notation $SO(10)$
or $SU(5)$.  Each vertex produces the same $\Delta (B-L)$ for all
fermions, since $B-L$ is gauged. However, $\Delta (B+L)$ is
different for standard and exotic fermions, because $(B+L)$
is just a global symmetry which can be broken consistently with
gauge symmetry, $E_6$.

It is still necessary to make the
assignment consistent with the lifetime of proton
and produce realistic fermion masses. This is achieved by introducing
discrete symmetries. We introduce a ${\cal Z}_2$ symmetry, under which,
\begin{equation}
\{ D_1, D_2, L_2, L_3\} \rightarrow - \{ D_1, D_2, L_2, L_3\}
\end{equation}
and all other particles remaining unchanged. The lagrangian
will then be given by,
\begin{eqnarray}
{\cal L} &=&  \lambda_{uu} Q^T u^c H_3 + \lambda_{dd} Q^T d^c H_2
+  \lambda_{DD} D_1^T D_2 S_2 + \lambda_{ee} L_1^T e^c H_2 \nonumber \\
&&+  \lambda_{EE} L_3^T L_2 S_2 + \lambda_{4}  L_1^T \nu^c H_3  + h.c.
\end{eqnarray}
The down quark mass matrix as well as the charged lepton
mass matrix are now diagonal. We assume the
mass of the exotic particles are in the range of 100 GeV, so that
this provides a rich $E_6$ phenomenology \cite{e6rev}. In addition, these
particles contribute to the diagrams which produce the $SU(2)_L$
anomaly during the electroweak phase transition. The field
$H_1$ now carries $(B-L)$ quantum number but does not
contribute to the mass of any particle and hence need not acquire
$vev$. $S_1$ carries $(B-L)$ quantum number and its $vev$
breaks the group $SU(2)_R \times U(1)_{B-L} \to U(1)_Y$
at a scale close to the GUT scale or at an intermediate scale.

There is still a problem with the neutrino masses, as it is the
case in most superstring inspired $E_6$-models. It can be
solved with one additional ${\bf 351}$ scalar representation, which
gives large majorana mass to the fields $\nu^c$ and $L_s$, which in
turn induce the see-saw mechanism. Another possibility is
for gravity to induce an effective term, which is similar
to the coupling of the fields in ${\bf 351}$ \cite{us2}.

The fields $H_2$ and $H_3$ carry zero baryon and lepton numbers.
The coupling of the scalar $S_2$ with coefficient $\lambda_{DD}$
requires this field to have $B=L=0$, but the coupling with
coefficient $\lambda_{EE}$ requires $B=L=1$. These two
couplings together will give rise to baryon and lepton number
violation, which has to be prevented. However, since these fields
do not couple to the ordinary particles directly, there is no danger
of proton decay. But their interactions can wash out any baryon
asymmetry of the universe, which must be suppressed.

These couplings can give rise to baryon number violation through
decays of $S_2$, shown in fig. 1, and through scattering
processes, $$D_1 + D_2 \to L_2 + L_3$$ shown in fig. 2.
These processes must be slow enough so that they are not in equilibrium
during the electroweak phase transition and destroy the baryon
asymmetry of the universe. In other words, the decay rates of
the field $S_2 \to L_2 + L_3$ and $D_1 + D_2$, as well as the
scattering rates for $ L_2 + L_3 \to D_1 + D_2$ at temperature
$T = M_S$ must be slower than the expansion rate of the universe,
\begin{equation}
\{ {\lambda_{EE}^2 \over {16 \pi}} M_S ,
{\lambda_{DD}^2 \over {16 \pi}} M_S,
\lambda_{EE} \lambda_{DD} {T^5 \over {(T^2 + M_S^2)^2}} \}
\leq 1.7 \sqrt{g_*} {T^2 \over M_{Pl}}
\end{equation}
All these conditions can be satisfied provided,
\begin{equation}
\lambda_{EE} \leq 10^{-5} \hskip .3in {\rm and} \hskip .3in
\lambda_{DD} \leq 10^{-5},
\end{equation}
which implies an intermediate symmetry breaking scale,
$M_S \sim \langle S_2 \rangle \geq 10^7$ GeV. So far we assigned
$B$ and $L$ quantum numbers to the exotic particles and discussed
the conditions under which the model is not ruled out by experiments.
The above conditions guarantee the consistency of the model but
they are not unique; alternative conditions consistent
with our $B$ and $L$ assignment are also possible.

As mentioned at the beginning of this article, in standard GUTs
(without any extra fermions beyond the standard model),
the operator analysis \cite{wein}, reveals
that all baryon number violating lowest dimensional operators, invariant
under the standard model gauge group, can be
written in the forms ${\bf q_L^T q_L q_L^T l_L; q_L^T q_L q_R^T l_R;
q_R^T q_R q_L^T l_L}$ and ${\bf q_R^T q_R l_R^T l_R}$.
This implies that there are no $SU(3)_c \times SU(2)_L \times U(1)_Y$
invariant operators with non-zero $(B-L)$. For this reason
most of the GUTs articles discuss only $(B-L)$ conserving proton decays.
$(B-L)$ violating proton decays are allowed only in some complicated
scenarios with higher dimensional operators \cite{pss}.
This restriction is not present in the $E_6$ model discussed.
In the rest of the paper we classify the new
operators which violate $(B-L)$ and a new scenario for baryogenesis.

In addition to the operators of the form mentioned above we can write
the following operators including
the exotic particles
which are invariant under the standard model,

\noindent
\vbox{
\begin{eqnarray}
{\bf Q \: Q \: Q \: L_2} \hskip .2in [1] & \hskip .2in
{\bf Q \: Q \: D_1 \: \nu^c } \hskip .2in [1] \hskip .2in &
{\bf Q \: Q \: D_1 \: L   } \hskip .2in [0] \nonumber \\
{\bf Q \: Q \: \overline{D_2} \: \overline{\nu^c} }
\hskip .2in [-1]& \hskip .2in
{\bf Q \: Q \: \overline{D_2} \: \overline{L}   }
\hskip .2in [0]\hskip .2in &
{\bf Q \: D_1 \: \overline{u^c} \: \overline{L_3} }
\hskip .2in [0]\nonumber \\
{\bf Q \: \overline{D_2} \: \overline{u^c} \: L_2}
\hskip .2in [0]& \hskip .2in
{\bf Q \: \overline{D_2} \: \overline{u^c} \: L_1  }
\hskip .2in [0]\hskip .2in &
{\bf Q \: \overline{u^c} \: \overline{d^c} L_2}
\hskip .2in [1] \nonumber \\
{\bf \overline{u^c} \: \overline{d^c} \: \overline{d^c} \: \overline{L}}
\hskip .2in [1] & \hskip .2in
{\bf \overline{u^c} \: \overline{d^c} \: \overline{d^c} \: \overline{\nu^c}}
\hskip .2in [0] \hskip .2in &
{\bf \overline{u^c} \: \overline{d^c} \: D_1 \: \nu^c}
\hskip .2in [1] \nonumber \\
{\bf \overline{u^c} \: \overline{d^c} \: D_1 \: \nu^c}
\hskip .2in [1] & \hskip .2in
{\bf \overline{u^c} \: \overline{d^c} \: \overline{D_2} \: \overline{\nu^c}}
\hskip .2in [-1] \hskip .2in &
{\bf \overline{u^c} \: \overline{d^c} \: \overline{D_2} \: \overline{L}}
\hskip .2in [0] \nonumber \\
{\bf \overline{u^c} \: {D_1} \: {D_1} \: \overline{\nu^c} }
\hskip .2in [-2] & \hskip .2in
{\bf \overline{u^c} \: {D_1} \: {D_1} \: \overline{L} }
\hskip .2in [-1] \hskip .2in &
{\bf \overline{u^c} \: \overline{D_2} \: D_1 \: \nu^c}
\hskip .2in [0] \nonumber \\
{\bf \overline{u^c} \: \overline{D_2} \: D_1 \: L}
\hskip .2in [-1] & \hskip .2in
{\bf \overline{u^c} \: \overline{D_2} \: \overline{D_2} \: \overline{L}}
\hskip .2in [-1] \hskip .2in &
{\bf \overline{u^c} \: \overline{D_2} \: \overline{D_2} \: \overline{\nu^c}}
\hskip .2in [-2] \nonumber
\end{eqnarray}}
\noindent
with the transportation sign being understood in the first and third
fields. The $(B-L)$ quantum numbers for the various operators are
given within the parentheses. It is now obvious that identifying the
$(B-L)$ quantum number with the generator of $E_6$ (as in the
$SO(10)$ GUT) we are allowed to have baryon and $(B-L)$ number violating
operator with only four fermions.

This allows new scenarios for baryogenesis.
We demonstrate it with a specific example.
Any operator of the form $(\overline{\psi^c_{iL}} \psi_{jL})
(\overline{\psi^c_{kR}} \psi_{lR})$
can be Fierz transformed to $(\overline{\psi^c_{iL}} \gamma^\mu \psi_{lR})
(\overline{\psi^c_{kR}} \gamma^\mu \psi_{jL})$.
In $E_6$ the gauge couplings,
$g (\overline{\psi^c_{iL}} \gamma^\mu \psi_{lR}) X_1^\mu$ and
$g (\overline{\psi^c_{kR}} \gamma^\mu \psi_{jL}) X_2^\mu$ are
present. The pair $X_1 X_2$ and $\overline{X_1} \:\overline{X_2}$
are then singlets under the $G_{std}$ but carry $(B-L) \neq 0$. They
can scatter, in general, by coupling to another scalar. The
net effect is that the scattering process  $X_1 + X_2 \to
\overline{X_1} + \overline{X_2}$ can generate unequal densities
of $X_1 X_2$ and $\overline{X_1} \:\overline{X_2}$ scattering states.
This occurs provided $CP$ is violated in the couplings of the
$X_1 X_2 \to {\rm scalar} \to \overline{X_1} \:\overline{X_2}$,
which is not forbidden in these theories. The densities of fermions
$(\overline{\psi^c_{iL}} \psi_{jL})
(\overline{\psi^c_{kR}} \psi_{lR})$ belongs to the operator classified
in the previous paragraph.

In summary, we pointed out that the baryon and lepton numbers of
the exotic particles of the $E_6$ model can have strong implications
to the low energy physics. We propose a new assignment of $B$ and $L$
quantum numbers for the exotic particles with the following features,
\begin{itemize}
\item[(1)] baryon and lepton number violating processes do not occur
at the electroweak phase transition, because the total $SU(2)_L$
anomaly for the $(B+L)$ quantum number is zero.
\item[(2)] baryon and lepton numbers are no longer global symmetries
of the low energy lagrangian.
\item[(3)] there are dimension six operators which violates
$(B-L)$ producing new $(B-L)$ violating processes, which
suggest a new scenario for the generation of the
baryon asymmetry in the universe.
\end{itemize}

\vskip .3in

{\bf Acknowledgement} We would like to thank  V. Kiselev, H. Nakano
and G. Zoupanos for discussions. U.~S. and H.~S. would like to acknowledge
fellowships from the Alexander von Humboldt Foundation and
hospitality from Institut f\"{u}r Physik, Universit\"{a}t Dortmund,
during their research stay in Germany. The financial support of DFG is
gratefully acknowledged.

\end{document}